# Highly efficient polaritonic light emitting diodes with angle-independent narrowband emission

Andreas Mischok[1,2]*, Sabina Hillebrandt[1,2], Seonil Kwon[1,2], Malte C. Gather[1,2]*

[1]Organic Semiconductor Centre, SUPA, School of Physics and Astronomy, University of St Andrews, North Haugh, St Andrews KY16 9SS, United Kingdom

[2]Humboldt Centre for Nano- and Biophotonics, Department of Chemistry, University of Cologne, Greinstr. 4-6, 50939 Köln, Germany

* andreas.mischok@uni-koeln.de, malte.gather@uni-koeln.de

**Abstract**

**Angle-independent, narrowband emission is required for many optoelectronic devices, ranging from high-definition displays to sensors. However, emerging materials for electroluminescent devices, such as organics and perovskites, show spectrally broad emission due to intrinsic disorder. Coupling this emission to an optical resonance reduces the linewidth, but at the cost of inheriting the severe angular dispersion of the resonator. Strongly coupling a dispersionless exciton state to a narrowband optical microcavity could overcome this issue; however, electrically pumped emission from the resulting polaritons has been hampered by poor efficiencies. Here, we present a universal concept for polariton-based emission from state-of-the-art organic LEDs (OLEDs) by introducing an assistant strong coupling layer, thus avoiding quenching-induced efficiency losses. We realize red- and green-emitting, narrowband (FWHM <20 nm) and spectrally tuneable polaritonic OLEDs with up to 10% external quantum efficiency and high luminance (>20,000 cd m$^{-2}$ at 5 V). Optimizing cavity detuning and coupling strength allows to achieve emission with ultralow-dispersion (<10 nm spectral shift at 60° tilt). These results have significant implications for on-demand polariton emission and demonstrate the practical relevance of strong light-matter coupling for next-generation optoelectronics, particularly display technology.**

**Introduction**

When a material with a strong excitonic resonance is inserted into a high-quality microcavity (MC), strong coupling and hybridisation of light and matter can occur and lead to the formation of exciton-polariton quasiparticles. Polaritons inherit a number of advantageous characteristics from their constituents, e.g., they show high interaction strength of excitons and are readily observable as emitted photons. Thus, polaritons offer a highly attractive platform for various applications[1], e.g. transistors[2,3], photodiodes[4–6], single-photon sources[7,8] and detectors[9], and polariton lasers acting as a Bose-Einstein condensate of polaritons[10–16]. However, in order to realise this potential, there is a need for a readily available polariton-based electroluminescent device that works under ambient conditions, i.e. a polariton

LED. Attempts to use GaAs to realize such LEDs have had limited success as the weak exciton binding in GaAs complicates operation at non-cryogenic temperatures[17].

Organic semiconductors offer large exciton binding energies and high oscillator strengths, thus polaritons are stable in these materials, even at room temperature[18]. Driving MC-based organic LEDs (OLEDs) into the strong coupling regime has indeed allowed the realization of room-temperature polariton LEDs[17,19,28,20–27]. However, so far, these polariton OLEDs have shown very low external quantum efficiency (EQE) and low brightness because strongly absorbing bulk materials had to be used as the emission layer (EML) to obtain a strong coupling effect. Such EMLs often have low photoluminescence quantum yields compared to the doped EMLs used in state-of-the-art OLEDs, all of which operate in the weak coupling regime[29]. Coupling inorganic and organic resonators[30,31] and employing radiative pumping of polaritons[32–35] have been proposed to improve the efficiency of OLEDs. Nevertheless, the highest EQEs reported so far for a single-cavity polariton OLED and a coupled weak-strong cavity architecture are only 0.2 %[28] and 1.2 %[35], respectively. In comparison, existing weakly coupled MC OLEDs readily achieve EQEs above 20%[36–39].

In this work, we demonstrate the realization of highly efficient, electrically driven polariton generation and light emission from both red- and green-emitting polaritonic OLEDs (POLEDs). To realize efficient operation, we add an assistant strong coupling layer into the second field maximum of a MC OLED that uses a doped EML with high luminescence quantum yield. The resulting devices retain more than 70 % of the EQE achieved by weakly coupled reference OLEDs and, at 5 V forward bias, reach luminance values of over 20,000 cd m$^{-2}$ in the red and 100,000 cd m$^{-2}$ in the green. We confirm that the characteristics of our devices are dominated by polariton formation, with emission exclusively emanating from the lower polariton branch (LPB). Varying the thicknesses of the cavity and strong coupling layer enables us to tune the emission characteristics and to enter the ultra-strong coupling regime without major losses in performance.

In addition to offering a strategy to realise polariton emission for a variety of emerging quantum technologies, we find that our POLEDs offer an attractive route towards the development of OLED displays with viewing-angle-independent emission colour and improved colour saturation. Many organic materials and the resulting OLEDs show relatively broad emission spectra due to the intrinsic disorder in these materials. Introducing these OLEDs into a weakly coupled MC leads to more narrowband emission and can improve colour saturation, but generally results in strong angular dispersion[36,40,41]. This dispersion manifests as a clearly visible change in the emission colour with increasing viewing angle and is problematic for OLED displays[42] and for emerging applications of OLEDs in biosensing and imaging[43]. Angular dispersion in MC OLEDs can be managed by the use of external[36] or internal[44,45] scattering elements or other designs[46,47] that mix the different angular components; however, these approaches substantially increase the spectral bandwidth and thus reduce colour purity. In contrast, we show here that the hybridised light-matter polaritons in our POLEDs can be tuned to exhibit exciton-like and thus flat angular dispersion, which eliminates any noticeable colour change with angle while maintaining the high colour purity and efficiency of a MC OLED. Thus, beyond their use in polaritonics, the POLEDs presented in this work offer highly desirable properties for display technology and other emerging applications that require narrowband and angle-independent light sources.

**Results and Discussion**

**Realization of polaritonic OLEDs**

We based the development of our POLED on an efficient phosphorescent MC OLED[39,48,49] that operates in the weak coupling regime and that serves as a reference device for our work. This reference OLED consists of a thin Ag bottom contact (25 nm) and a thick Ag top contact that form a second-order bottom-emitting MC and sandwich a thick p-doped hole transport layer (HTL), electron-blocking layer (EBL), emitter layer (EML), hole-blocking layer (HBL), and n-doped electron transport layer (ETL) as summarized in Fig. 1a and Supporting Information Fig. S1. Transfer matrix (TM) simulations were performed to optimize light outcoupling at the emission wavelength of the Ir(MDQ)$_2$(acac)-based EML. The reference OLED generates a narrowband emission spectrum that can be tuned in wavelength by adjusting the thickness of the charge transport layers. The use of doped charge transport layers means that these changes in layer thickness have negligible impact on the operating voltage of the final device and also ensure low-voltage operation.

Strong light-matter coupling occurs when the interaction between photons and excitons in the system becomes stronger than the loss rate of either particle. Due to the weakly allowed triplet-to-singlet transition of state-of-the-art phosphorescent emitters like Ir(MDQ)$_2$(acac) and because their concentration in the EML is relatively low (10 wt% for our devices), these emitters on their own do not exhibit sufficient interaction with cavity photons for strong coupling to occur. Instead, we hypothesize that the addition of an absorber with an absorption spectrum complimentary to the emission of Ir(MDQ)$_2$(acac) in the second field maximum of the cavity can facilitate strong coupling without jeopardizing OLED performance. This strategy splits up the processes of electroluminescence (by Ir(MDQ)$_2$(acac)) and polariton formation (predominantly within the absorber), and thus realizes both processes efficiently in a single monolithic device stack.

Boron-substituted phthalocyanines are commonly used in non-fullerene organic solar cells and have recently been shown to induce strong coupling, even when present as a relatively thin layer, due to their strong absorption and low Stokes shift[5]. Cl$_6$SubPc provides nearly perfect complimentary absorption to the Ir(MDQ)$_2$(acac) emission from the EML (Fig. 1c). As the position of the highest occupied molecular orbital (HOMO) of Cl$_6$SubPc at $-5.6$ eV[50] is close to that of the Spiro-TTB-based HTL at $-5.3$ eV, we expect that a thin layer of Cl$_6$SubPc can be inserted into the HTL without creating a significant energy barrier to hole transport. To further reduce the impact of the addition of $Cl_6SubPc$ on electrical performance and avoid crystallization of $Cl_6SubPc$[51,52], we co-evaporated $Cl_6SubPc$ with Spiro-TTB at a volumetric ratio of 1:1.

To realize our POLED, a 24 nm-thick layer of the Spiro-TTB:$Cl_6SubPc$ blend was positioned within the HTL of the OLED, so that the layer was located exactly in the field maximum of the cavity mode at the wavelength of the $Cl_6SubPc$ exciton (590 nm, Fig. 1b). Simultaneously, the thickness of the HTL was reduced by 32 nm to maintain the same overall optical thickness as the reference MC OLEDs (see Supporting Information Tables S1 and S2 for the layer thicknesses of all devices in this study).

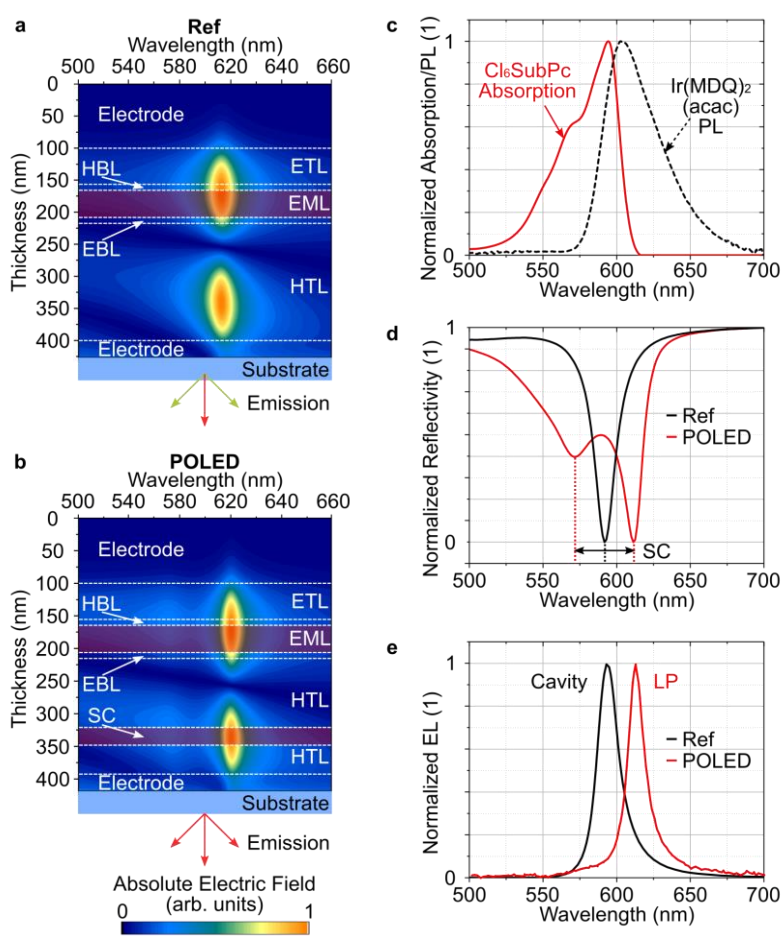

**Figure 1. Concept and initial demonstration of the polaritonic OLED (POLED) based on an assistant strong coupling layer. a**, Structure and electric field distribution of the reference OLED comprised of two reflective Ag electrodes forming a microcavity, Spiro-TTB:$F_6$TCNNQ at 4 wt% as the p-doped hole-transport layer (HTL), NPB as the electron-blocking layer (EBL), NPB doped with Ir(MDQ)$_2$(acac) at 10 wt% as the emission layer (EML), BAlq as the hole-blocking layer (HBL), and BPhen:Cs at 3 wt% as the n-doped electron-transport layer (ETL). **b,** Structure and electric field distribution of the POLED with a 24 nm-thick assistant strong coupling layer (SC) of the absorber $Cl_6SubPc$ (blended 1:1 with SpiroTTB) inserted into the HTL at the location of peak optical field strength.

**c**, Absorption spectrum of Cl$_6$SubPc (red solid line) and photoluminescence (PL) spectrum of Ir(MDQ)$_2$(acac) (black dashed line). **d**, Reflectivity spectra at 20° angle of incidence for a reference microcavity OLED without the strong coupling layer (black) and a POLED (red). Strong coupling of cavity photons and Cl$_6$SubPc excitons leads to splitting of the resonance peak in the POLED reflectivity spectrum, as indicated by the double arrow. **e,** Corresponding electroluminescence (EL) spectra of both devices at a viewing angle of 20°.

While the reflectivity spectrum of the reference OLED has a single characteristic dip associated with the cavity mode, the reflectivity spectrum of the POLED shows a distinct splitting into an upper and lower polariton (Fig. 1d), which provides evidence of strong coupling between cavity photons and material excitons. The presence of strong coupling was further confirmed by a red shift of the electroluminescence, which was aligned with the cavity mode for the reference OLED but emanated exclusively from the LPB in the POLED (Fig. 1e, there is a weak shoulder visible in the emission spectrum at 575 nm, originating from emission outside the cavity, see Supporting Information Fig. S2). Despite the different nature of their emission processes, the reference OLED and the POLED both have narrowband emission spectra, with Lorentzian linewidths of 16 nm; much sharper than the photoluminescence spectrum of Ir(MDQ)$_2$(acac) (Fig. 1c) and the emission spectrum of a conventional, non-cavity-based Ir(MDQ)$_2$(acac) OLED (Supporting Information Fig. S7). To assess the contributions of Ir(MDQ)$_2$(acac) and Cl$_6$SubPc to strong coupling, we have a performed careful analysis of the absorption spectrum of the individual layers combined with transfer matrix simulation of the intracavity absorption (Supporting Information Fig. S3 and Supporting Note 1).

**Variation of cavity detuning**

Next, we explored how varying the optical thickness and thus detuning the cavity mode relative to the exciton energy affects polariton formation and angular dispersion. Figure 2 summarizes the transfer-matrix modelled and experimentally measured angle-resolved reflectivity spectra of weakly coupled reference OLEDs with a range of cavity thicknesses (Ref-A to Ref-D) and of the corresponding strongly coupled POLEDs (P-A to P-D). The reflectivity spectra of the POLEDs show clear anti-crossing of the upper and lower polariton branches, confirming the presence of strong coupling in the POLEDs over the entire range of detuning investigated.

To confirm our interpretation and to measure the coupling strength in the POLEDs, we compared our observations to the predictions of the coupled oscillator (CO) Hamiltonian[53]

$$\begin{pmatrix} E_{\text{Cav}}(\theta) & G \\ G & E_X \end{pmatrix} \begin{pmatrix} F_{\text{Cav}} \\ F_X \end{pmatrix} = E_{\text{Pol}}(\theta) \begin{pmatrix} F_{\text{Cav}} \\ F_X \end{pmatrix},$$

where $E_{\text{Cav}}(\theta)$ and $E_X$ are the cavity and exciton energies, $G$ is the Rabi-splitting of lower and upper polariton, and $F_{\text{Cav}}$ and $F_X$ are the cavity photon fraction and exciton fraction of the polariton, respectively, $E_{\text{Pol}}(\theta)$ is the polariton energy and $\theta$ is the outcoupling angle. The energy of the cavity mode $E_{\text{Cav}}(\theta)$ was extracted from the reflectivity of the reference OLEDs and the exciton energy was determined from the Cl$_6$SubPc absorption spectrum. (As the excitons of Cl$_6$SubPc and Ir(MDQ)$_2$(acac) are overlapping in energy, they were not modelled separately to avoid unnecessary ambiguity in the model.) We then varied the coupling strength in the CO Hamiltonian until the predicted polariton energies agreed with the features in the reflectivity spectra of each POLED. For all POLEDs, we obtained excellent agreement for a Rabi-splitting energy of $G = 140$ meV (*c.f.*, solid lines in Fig. 2). The analysis with the CO Hamiltonian also shows that the exciton fraction in the LPB is highest for the POLEDs with the highest $E_{\text{Cav}}$ and is larger for larger angles (Supporting Information Fig. S4). Consequently, we observe a predominantly flat, exciton-like dispersion of the LPB in Devices P-A and P-B. As a rule, high coupling strengths and a blue-shifted microcavity with respect to the exciton line appear to lead to particularly flat angular dispersion of the LPB.

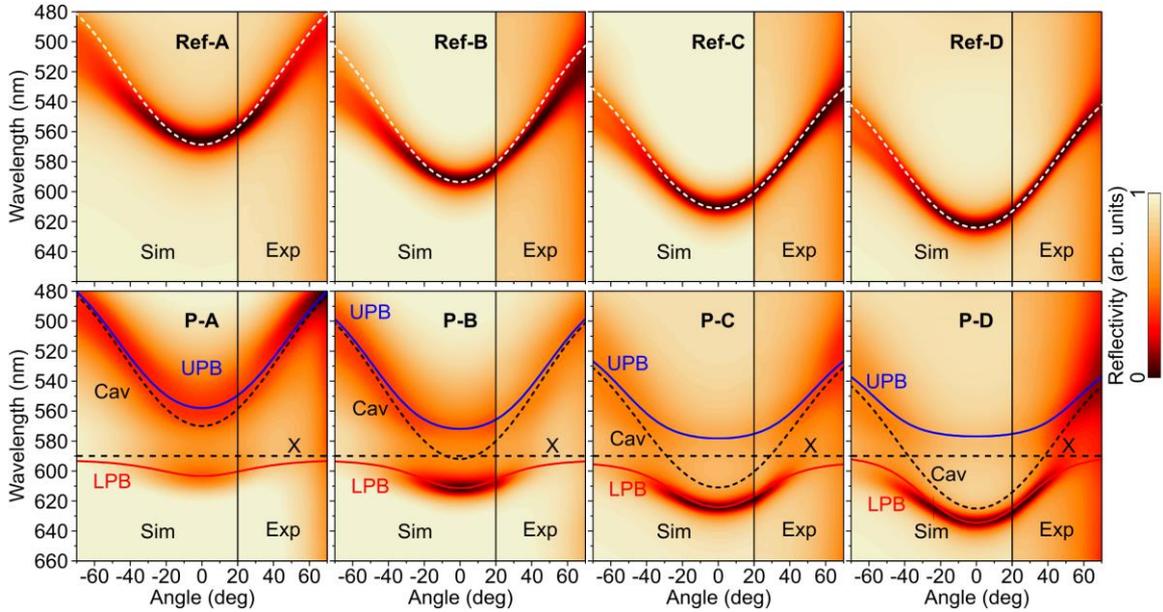

**Figure 2. Polariton formation and tuning.** Simulated (Sim, left side) and measured (Exp, right side) reflectivity spectra of weakly coupled reference OLEDs (top row) and of POLEDs with the strong coupling layer (bottom row) on a false-colour scale. Cavity thickness increases from left to right; reference OLEDs and POLEDs in the same column have the same total optical thickness. Dashed white and black lines indicate the calculated position of the cavity mode and exciton (X) in the reference OLEDs and POLEDs, respectively. The solid blue and red lines in the bottom row show the position of the upper and lower polariton branches, respectively, as obtained from the coupled oscillator model.

**Electrical performance**

Figure 3 shows the current-voltage-luminance characteristics and the EQE versus luminance data for the reference OLEDs Ref-A to Ref-D and the POLEDs P-A to P-D. Comparing the reference and polaritonic OLEDs, the most important observations are: (i) both sets of devices reach high luminance values above 20,000 cd m$^{-2}$ at a low operating voltage of 5 V and show emission at wavelengths between 570 and 640 nm. (ii) The introduction of the absorbing strong-coupling layer in the POLED does not lead to a significant change in turn-on voltage but causes a reduction in current density (by approx. 10-50 % depending on device and applied voltage). (iii) The EQE increases from Ref-A to Ref-D and from P-A to P-D due to an increase in outcoupling efficiency for more red-tuned devices. (iv) The POLEDs reach approximately 70 % of the EQE of the corresponding reference OLEDs, peaking at an EQE of 9.7 % for P-D. (v) The roll-off in efficiency[54] at high luminance is not much affected by the strong coupling layer (e.g., critical current density $J_{90\%}$ = 21 mA cm$^{-2}$ for P-D versus 15 mA cm$^{-2}$ for Ref-D), indicating that the polariton conversion process in the devices operates far from saturation. (vi) The POLEDs retain the strong directional emission of the microcavity devices, as evidenced by the polar plots of both device types in Fig. 3e,f.

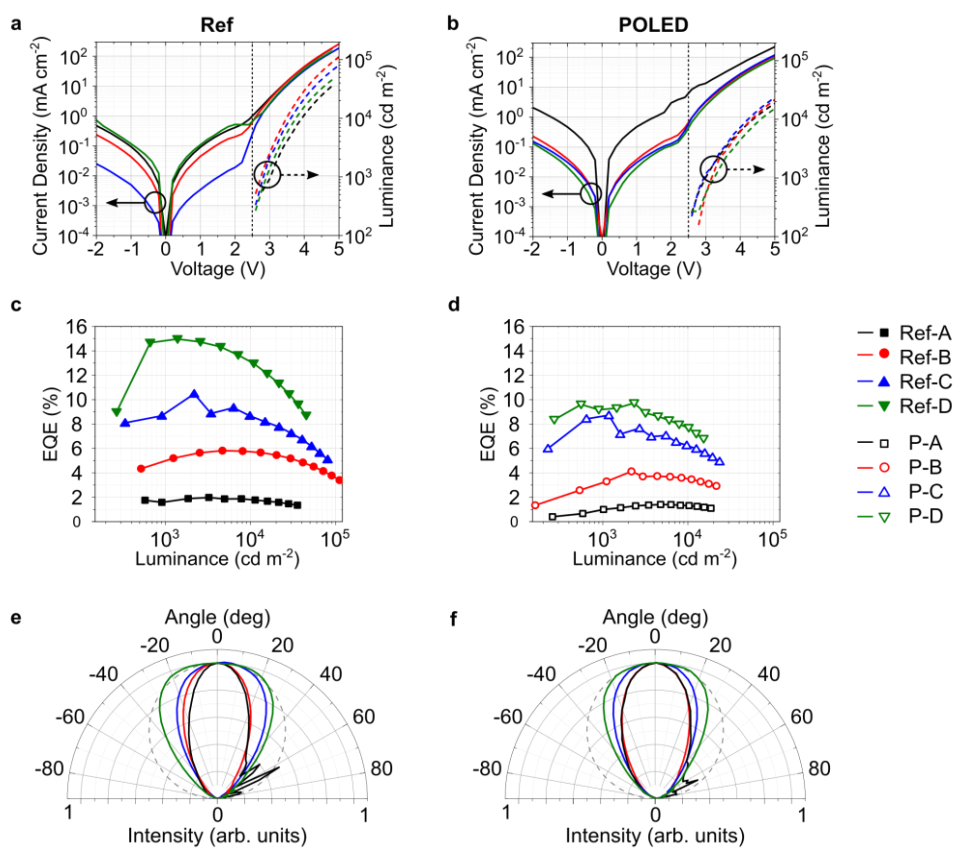

**Figure 3. Device performance. a, b**, Current density and luminance over voltage for the reference OLEDs (a) and the POLEDs (b). All devices turn-on around 2.5 V (dashed vertical line) and reach a luminance of over 10,000 cd/m$^2$ at 5 V, irrespective of whether the strong coupling layer is present in the device or not. **c, d**, External quantum efficiency over luminance for the reference OLEDs (c) and the POLEDs (d), respectively. **e, f**, Angle resolved emission intensity of reference (e) and polaritonic (f) OLEDs, showing directional emission for all devices. Dashed lines indicate the profile of a Lambertian emitter. The EQE values in **c** and **d** take the actual, non-Lambertian angular emission characteristics of each device into account.

**Angular colour stability and flexible POLEDs**

The transformation of cavity photons into polaritons strongly alters the angular dispersion of the light emitted by our devices. In particular, as the dispersion of the LPB becomes more exciton-like, its wavelength shift with angle is greatly reduced. As the emission occurs exclusively from the LPB in our POLEDs, this leads to greatly improved angular colour stability of the emission. Figure 4a,b compares the angle-resolved electroluminescence of the weakly coupled MC OLED Ref-C to the POLEDs P-A and P-B, which all emit at similar wavelengths at 0 ° observation angle (see Supporting Information Fig. S5 and S6 for angle-resolved electroluminescence spectra of all devices in this study). While the colour of the reference OLED visually changes from red to yellow when the device is tilted, no change in colour is noticeable for the POLEDs. The peak emission wavelength of the reference OLED blue-shifts by 45 nm, from 612 nm at 0 ° to 567 nm at 60 °. In contrast, the POLEDs only show a wavelength shift of 8 nm (P-A, from 607 nm at 0 °) and 15 nm (P-B, from 617 nm), respectively. The residual wavelength shifts of the POLEDs over the 0 to 60 ° cone are thus smaller than or comparable to the FWHM of their respective emission spectra (P-A 18 nm FWHM, P-B 12 nm). For comparison, Figure S7 shows the much broader emission spectrum (70 nm FWHM) of a non-MC OLED with analogous structures but with a transparent ITO bottom contact.

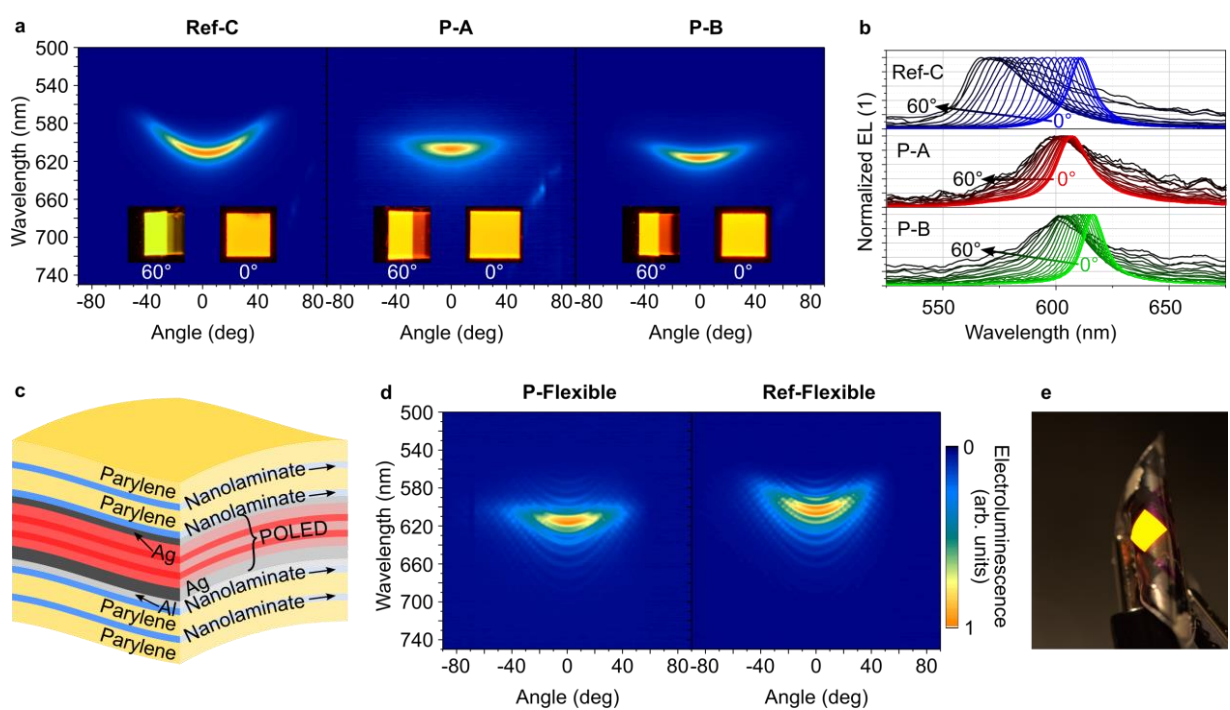

**Figure 4. Polaritonic OLEDs provide angle-independent emission spectra and allow mechanically flexible devices. a**, Angle-resolved electroluminescence spectra of the reference OLED Ref-C and POLEDs P-A and P-B on a false-colour scale. The insets show photographs of the emission of each OLED at observation angles of 0 ° and 60 °, respectively. **b**, Normalized electroluminescence spectra of the devices in **a** for angles between 0 and 60 °. All spectra were recorded at a constant current

density of 25 mA cm$^{-2}$. **c**, Device structure of the flexible POLED sandwiched by flexible thin-film encapsulation without a supporting substrate. The encapsulation is based on layers of parylene-C and a metal oxide nanolaminate. **d**, Angle-resolved electroluminescence spectrum for the flexible POLED and for a flexible reference OLED (also using a MC, but without a strong coupling layer). **e**, Photograph of emission from a rolled up, flexible POLED showing no change in emission colour, irrespective of the viewing and bending angle.

These results demonstrate how strong light-matter coupling can be exploited to realize efficient, narrowband, and colour-stable OLED emission, avoiding the need for electrodes made from expensive and often brittle transparent conductive oxides such as ITO. This opens the opportunity to realize flexible OLED displays with unprecedented colour saturation and colour stability. To explore this possibility, we next developed a substrateless and flexible POLED (Fig. 4c). In this device, a thin-film encapsulation stack—based on layers of parylene-C and $Al_2O_3/ZrO_2$ nanolaminates—acts as both a moisture and oxygen barrier and a quasi-substrate[55], thus reducing the total thickness of the device to below 30 μm. The resulting flexible POLED exhibits narrowband (FWHM 25 nm) and virtually angle-independent emission spectra (13 nm shift from 0 to 60 ° viewing angle) whereas a flexible reference MC OLED shows the expected large angular dispersion (Fig. 4d). Due to interaction between the OLED microcavity and the thin film encapsulation stack, the emission spectrum of the flexible POLED is slightly broader than the spectra of our earlier devices on rigid substrates. In addition, weak parabolic cavity modes originating from the parylene-C layers are visible in the spectrum; however, this does not diminish the visually perceived colour stability of the device. Even under extreme bending (radius of curvature <5 mm), when an observer sees the emissive surface of the device under a range of angles, no variation in emission colour is visible (Fig. 4e).

**Tuning of the coupling strength**

As we introduce strong coupling in our POLEDs in a manner that separates polariton formation from OLED emission, the coupling strength can be tuned without significantly changing the electrical properties of the devices. Figure 5a-c shows the simulated and measured polariton dispersion for a series of devices with strong coupling layers of increasing thickness. The Rabi-splitting energies in these devices increase from 140 meV for a 24 nm SpiroTTB:Cl$_6$SubPc layer, to 190 meV for 36 nm and to 240 meV for 48 nm, as determined by CO modelling. This demonstrates direct tuning of the coupling strength without majorly impacting the performance of the device compared to the respective reference OLEDs (Supporting Information Fig. S8). As the emission still occurs exclusively from the LPB (Supporting Information Fig. S9), we conclude that the characteristics of the OLED are dominated by polariton formation, and thus they can be tuned over a wide range by simply adjusting the thickness of the strong coupling layer.

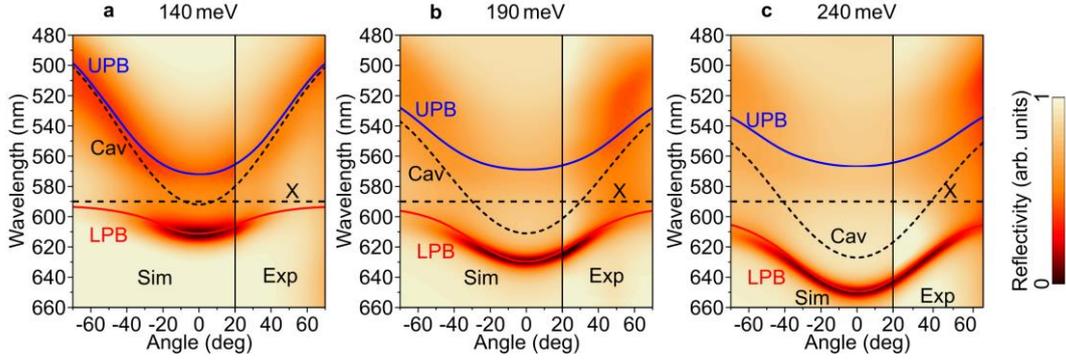

**Figure 5. Variation of coupling strength in red polariton OLEDs.** Simulated (Sim, left side) and measured (Exp, right side) reflectivity spectra for POLEDs with strong coupling layers of different thickness, resulting in increasing Rabi-splitting. **a**, 24 nm thickness, G = 140 meV, **b**, 36 nm thickness, G = 190 meV (b), and **c**, 48 nm thickness, G = 240 meV. The dashed black lines indicate the calculated position of the cavity modes and the exciton line (X). The solid blue and red lines show the position of the upper and lower polariton branches, respectively, as obtained from the coupled oscillator model.

**Fully optimized polaritonic OLEDs with green emission**

The use of the assistant strong coupling layer is not limited to the specific combination of emitter and strong coupling materials presented above. To show the universality of the concept, we realized POLEDs comprising the phosphorescent green emitter Ir(ppy)$_2$(acac) combined with an assistant strong coupling layer of the coumarin dye C545T[28,53]. As in our original material system, this material pair exhibits complimentary absorption (C545T) and emission (Ir(ppy)$_2$(acac)) spectra.

Adapting a MC OLED previously reported in Ref. 56, we fabricated a set of reference MC OLEDs and POLEDs with these materials (Supporting Information Figs. S10 and S11 and Tables S4 and S5). We then performed the same simulations and measurements of angle-resolved reflectivity, angle-resolved EL and electrical performance as described above for the red-emitting devices. Again, we observed angle-independent polariton emission with high luminance (here, >100,000 cd m$^{-2}$ at 5 V) and a peak EQE of 10.2 % (Supporting Information Fig. S12 and S13). As before, the efficiency of the green POLEDs is broadly comparable to the efficiency of the corresponding references. However, the green-emitting POLEDs reach significantly lower current densities than the reference MC OLEDs (e.g., approx. 300 mA cm$^{-2}$ at 5 V versus 2000 mA cm$^{-2}$ for a corresponding weakly coupled reference OLED), suggesting that adding a bulk C545T layer with low hole mobility introduces a charge imbalance. This illustrates the importance of choosing either an assistant strong coupling material with adequate charge mobility or using the blended layer approach we applied for red-emitting POLEDs. However, as the assistant strong coupling layer would generally be embedded inside the HTL or the ETL, i.e., in a monopolar and doped region of the device stack, we expect that suitable material combinations can be identified for most device architectures and charge transport layers.

As for the red-emitting POLEDs, the coupling strength of the green system can be tuned by varying the thickness of the strong coupling layer (Supporting Information Fig. S14). Here, we find that the combination of Ir(ppy)$_2$(acac) and C545T works particularly well even for higher coupling strengths. Combining the optical tuning through the cavity thickness and tuning of the coupling strength, we realized a fully optimized device that operates in the ultra-strong coupling (USC) regime and for which the polariton shows a high exciton fraction. Figure 6 summarizes the characteristics of this green USC POLED with an optimized stack (Supporting Information Table S6) that includes a 60 nm thick assistant strong coupling layer of C545T, which pushes the Rabi-splitting to 520 meV (i.e., a coupling strength of 10.3% of the bare exciton energy), and that creates highly favourable outcoupling conditions for light emission. This optimized device reached an EQE of 9.7 %, a luminance above 20,000 cd m$^{-2}$ at 5 V and a flat dispersion with a shift of <20 nm at 60 ° viewing angle. This represents an increase in EQE relative to the previous record for polariton OLEDs[28] by more than one order of magnitude. Further USC POLEDs with different detunings, along with a more detailed analysis of the optimized device are shown in Supporting Information Figs. S15-S17 and Supporting Note 2.

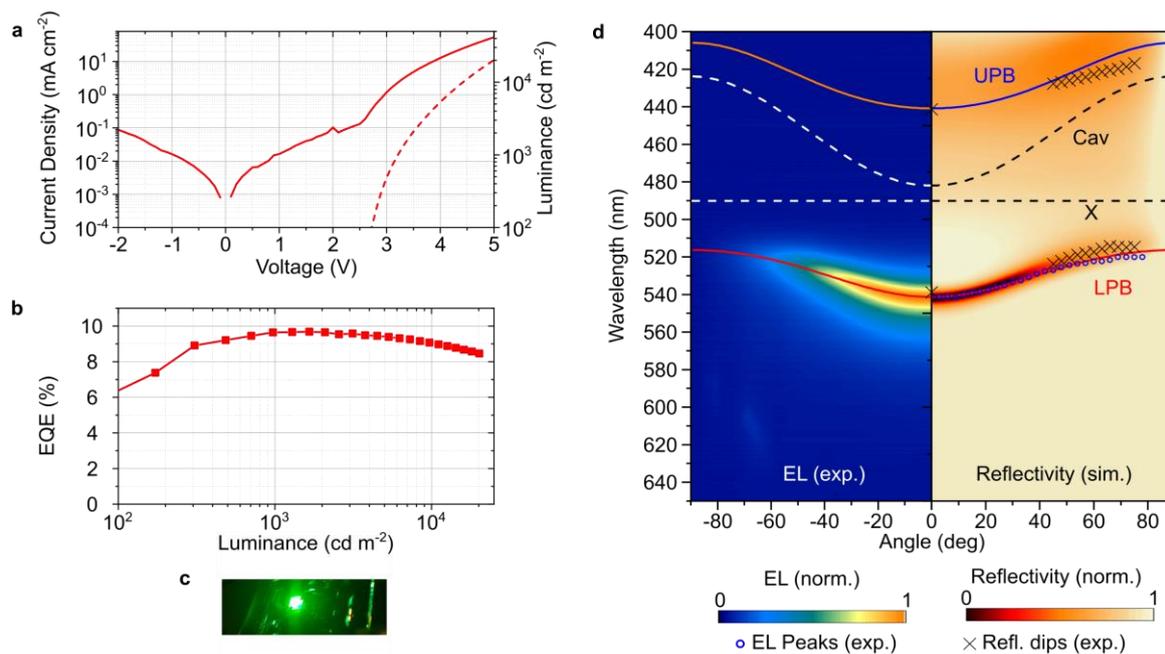

**Figure 6: Optimized green POLED in the ultra-strong coupling regime. a**, Current density-voltage-luminance characteristics, showing the device reaches a luminance of 20,000 cd m$^{-2}$ at 5 V in the presence of a thick assistant strong coupling layer. **b**, External quantum efficiency of the device, reaching a peak of 9.7% at ~1,500 cd m$^{-2}$. **c**, Photograph of POLED emission. **d**, Angle-resolved, measured electroluminescence (EL) intensity (left half) and simulated device reflectivity (right half) of the USC POLED. Crosses show positions of minima in measured reflectivity spectra and circles show peak position of the measured EL intensity. Solid lines show calculated polariton branches while dashed lines indicate the cavity mode and exciton energy.

**Discussion**

By adding a strong coupling layer in the second field maximum of a second order cavity device, a state-of-the-art phosphorescent OLED structure was transformed into a bright and highly efficient source of on-demand, electrically pumped polariton emission. Combining efficient Ir-based emitters with either $Cl_6SubPc$ in the red or C545T in the green yielded highly tunable polariton emission, with EQEs of 10 % and luminance levels over 20,000 cd m$^{-2}$ (red) and 100,000 cd m$^{-2}$ (green), respectively, at 5 V forward bias. In spite of the low absorption of the emissive layers used in our devices, the assistant strong coupling layer ensures that the overall device characteristics remain dominated by the strongly coupled cavity; at a sufficiently large coupling strength, even the ultra-strong coupling regime is reached without introducing major losses in device efficiency. In comparison to previously demonstrated record polariton OLEDs[28,35], this represents an increase in EQE of at least an order of magnitude, and an increase in luminance and luminous efficacy of at least two orders of magnitude, thus for the first time demonstrating that such devices can reach an application-relevant performance. Combining our concept with thin film encapsulation, we also realized—to the best of our knowledge—the first demonstration of a mechanically flexible polariton LED.

Earlier POLEDs were typically based on bulk emitter layers. These materials generally show low luminescence quantum yields and do not support mechanisms to harvest triplet states for electroluminescence. By decoupling light emission and polariton formation, we avoid the catastrophic losses in device efficiency that have plagued POLEDs so far, gain access to phosphorescent emitters, and thus increase POLED efficiency by an order of magnitude over the state of the art. Our realization of efficient red and green POLEDs indicates that the majority of currently used emitter materials can be employed as an emissive layer in our polariton devices, as long as they are combined with the appropriate assistant strong coupling layer. Given the enormous breadth of OLED structures known today, the strategy of using an assistant strong coupling layer therefore unlocks the entire visible and parts of the near IR spectrum for efficient polariton emission. In addition, this universality provides numerous possibilities to further increase device efficiency and thus reduce the remaining gap in EQE between POLEDs and conventional OLEDs, e.g., through the application of the latest generation phosphorescent and thermally activated delayed fluorescence emitters including emitter systems showing strongly horizontal dipole orientation[57–60], and the integration of capping layers and other light extraction modalities[61]. This will also reveal if the fundamental limitations to POLED efficiency differ from those for conventional OLEDs.

In the future, the POLEDs developed here could be adopted to use a short excited-state lifetime fluorescent emitter and a high luminescence quantum yield strong coupling layer. Such an architecture would provide a promising monolithic pathway to electrically pumped polariton lasing in organic systems. As with existing work on electrically pumped lasing, a further challenge is to realize a high-Q cavity in the presence of electrical contacts. Tamm-plasmon-polariton cavities which combine thin metallic contacts with high optical quality distributed Bragg reflectors represent a promising strategy to achieve this[62–64].

Beyond their relevance to lasers and fundamental studies of polariton phenomena, a number of features render the POLEDs developed in this work particularly attractive for display applications: most importantly, by appropriate tuning of the emissive lower polariton state, the angular dependence of the emission wavelength in these narrowband devices can be greatly reduced. This enables the emission of more saturated colours than possible with current OLED technology, without introducing a change in the perceived colour when looking at the device from different angles. In the past, changes in emission colour with viewing angle have prevented or complicated the use of MC OLEDs based on metallic thin film electrodes and have often necessitated the application of a transparent conductive oxide for one of the electrodes. However, the brittle nature of many conductive oxides and the scarcity of indium, which is a common component in these devices[65], makes a move to OLEDs based entirely on thin metal electrodes highly desirable, both from a cost perspective and to improve the robustness and reliability of future flexible displays. Furthermore, the integration of OLEDs on some types of driver backplanes, in particular those based on silicon CMOS chips, requires the use of top-emitting OLED architectures. For these architectures, metal-metal MC OLEDs are again preferred over devices using a transparent conductive oxide top-electrode, as deposition of the latter material onto the organic layers of the OLED stack can damage the organic material and thus degrade device performance. Lastly, the rearrangement of the exciton energies to polariton states that occurs in a POLED can reduce the effective optical bandgap of the system and enable tuning of the singlet energies, and this strategy may allow to enhance device efficiency through improved recycling of triplet states[5,66–69].

**Acknowledgments**

The authors acknowledge funding by the Volkswagen Foundation (No. 93404), the Leverhulme trust (RPG-2017-213), the European Research Council under the European Union Horizon 2020 Framework Programme (FP/2014-2020)/ERC Grant Agreement No. 640012 (ABLASE), and the Alexander von Humboldt Foundation (Humboldt Professorship to MCG). A.M. acknowledges further funding through an individual fellowship of the Deutsche Forschungsgemeinschaft (404587082) and from the European Union's Horizon 2020 research and innovation programme under Marie Skłodowska-Curie grant agreement No. 101023743 (PolDev).

**Author contributions**

A.M., S.H., and M.C.G. initiated the study and planned the experiments. A.M., S.K. and S.H. fabricated and measured the electrical performance of devices. S.H. and S.K. optimized the reference OLEDs. A.M. and S.K. fabricated and analysed the flexible devices. A.M. performed the angle-resolved reflectivity and ellipsometry measurements, device photography, and calculations using the CO and TM models. All authors analysed and discussed the data. A.M. and M.C.G. wrote the paper with contributions from all authors.

**Competing interests**

The authors declare no competing interests.

**Experimental methods**

*OLED materials and fabrication*: OLEDs were fabricated via thermal evaporation of organic and metal thin films at a base pressure of $1 \times 10^{-7}$ mbar (Angstrom EvoVac) onto 1.1 mm-thick glass substrates. The materials used were: Ag as a bottom and top contact, 2,2',7,7'-tetrakis(N,N'-di-p-methylphenylamino)-9,9'-spirobifluorene (Spiro-TTB) p-doped with 2,2'-(perfluoronaphthalene-2,6-diylidene)dimalononitrile (F6-TCNNQ) (4 wt%) as a hole transport layer; 2,3,9,10,16,17-hexachlorinated boron subphthalocyanine chloride ($Cl_6SubPc$) and 2,3,6,7-tetrahydro-1,1,7,7,-tetramethyl-1H,5H,11H-10-(2-benzothiazolyl)quinolizino-[9,9a,1gh]-coumarin (C545T) as strong coupling layers; N,N'-di(naphtalene-1-yl)-N,N'-diphenylbenzidine (NPB) or 1,1-Bis[(di-4-tolylamino)-phenyl]cyclohexane (TAPC) as an electron-blocking layer; NPB or tris(4-carbazoyl-9-ylphenyl)amine (TCTA) as an emitter host layer; bis(2-methyldibenzo[f,h] quinoxaline)(acetylacetonate)-iridium(III) ($Ir(MDQ)_2(acac)$, doping concentration 10 wt%) or bis[2-(2-pyridinyl-N)phenyl-C] (acetylacetonato)-iridium(III) ($Ir(ppy)_2(acac)$, doping concentration 10 wt%) as emitting dopands; bis-(2-methyl-8-chinolinolato)-(4-phenyl-phenolato)-aluminium(III) (BAlq) or 4,7-Diphenyl-1,10-phenanthroline (BPhen) as a hole-blocking layer; and BPhen doped with Cs as an electron transport layer. All organic materials were obtained from Lumtec in sublimed grade and used as received. Layer thicknesses were controlled in situ using calibrated quartz crystal microbalances (QCMs). Details of the device stack and layer thicknesses for each sample are provided in the Supporting Information (Tables S1–S5). Before fabrication, substrates were cleaned by ultrasonication in acetone, isopropyl alcohol and deionized water (10 min each), followed by $O_2$ plasma-ashing for 3 min. Evaporated devices were encapsulated without exposure to ambient air in a nitrogen atmosphere with a glass lid and moisture getter using UV-curable epoxy (Norland NOA68). The active area of the devices was 4.0 mm$^2$. Flexible devices based on thin-film encapsulation were prepared on cleaned glass carrier substrates, as described in Keum e*t al.*[55]. In brief, parylene-C (diX C, KISCO) and the thin-film oxide layers were deposited using a parylene coater (Labcoater 2, SCS) and an atomic layer deposition (ALD) reactor (Savannah S200, Ultratech), respectively, with both coaters connected to the evaporator in a common nitrogen filled glovebox. In order to accommodate the processing temperature of the ALD layers (80 °C), BPhen was replaced by the more stable 2,9-dinaphthalen-2-yl-4,7-diphenyl-1,10-phenanthroline (NBPhen) and a hole injection layer of 1 nm $MoO_3$ was added between the bottom contact and HTL[70]. Details of the flexible device stack are presented in Supporting Information Table S3. After processing, the devices were peeled off the carrier substrate to yield free-standing, flexible OLEDs.

*Device characterization:* Current density and luminance-voltage behaviour were analysed with a source measure unit (SMU, Keithley 2450) and a calibrated amplified Si photodiode (Thorlabs PDA100A) positioned at a fixed distance of 16.7 cm and connected to a digital multimeter (Keithley 2100). Angle-resolved spectra were measured using a goniometer setup built in-house[71] equipped with a fibre-coupled spectrometer (OceanOptics Maya LSL). External quantum efficiency (EQE) was calculated by taking into account the angular emission characteristics of the OLED. Angle-resolved reflectivity spectra were recorded using a varying angle spectroscopic ellipsometer which allowed reflectivity measurements for angles >20° (VASE, M2000, J.A. Woollam). The optical constants of all materials were determined using VASE and used to calculate their absorption profiles, electric field distribution, and OLED reflectivity spectra with a transfer matrix model. Polariton branches were calculated using a coupled oscillator model[53]. The photographs of the devices were taken using a digital single-lens reflex camera (Nikon D7100) with a macro lens (Sigma 105 mm F2.8 EX DG OS HSM).

**Data availability**

The research data supporting this publication can be accessed via the University of St Andrews repository at https://doi.org/<to be assigned upon acceptance>

# Highly efficient polaritonic light emitting diodes with angle-independent narrowband emission

Andreas Mischok[1,2]*, Sabina Hillebrandt[1,2], Seonil Kwon[1,2], Malte C. Gather[1,2]*

[1]Organic Semiconductor Centre, SUPA, School of Physics and Astronomy, University of St Andrews, North Haugh, St Andrews KY16 9SS, United Kingdom

[2]Humboldt Centre for Nano- and Biophotonics, Department of Chemistry, University of Cologne, Greinstr. 4-6, 50939 Köln, Germany

* andreas.mischok@uni-koeln.de, malte.gather@uni-koeln.de

**Table of Contents**

| Layer | Thickness (nm) | | | |
|---|---|---|---|---|
| | Ref-A | Ref-B | Ref-C | Ref-D |
| Ag | 100 | 100 | 100 | 100 |
| BPhen:Cs | **50** | **60** | **70** | **80** |
| BAlq | 10 | 10 | 10 | 10 |
| NPB:Ir(MDQ)$_2$(acac) | 40 | 40 | 40 | 40 |
| NPB | 10 | 10 | 10 | 10 |
| SpiroTTB:F$_6$TCNNQ | **197** | **202** | **207** | **212** |
| Ag | 25 | 25 | 25 | 25 |
| Substrate | 1.1E6 | 1.1E6 | 1.1E6 | 1.1E6 |

**Table S1. Device structure and layer thicknesses of red reference OLEDs.** Nominal layer thicknesses for red reference OLEDs comprising the emitter Ir(MDQ)$_2$(acac) with variations in cavity thickness (by changing the thickness of the SpiroTTB:F$_6$TCNNQ HTL and of the BPhen:Cs ETL). Due to tooling variations and chamber positioning, the real film thicknesses can differ by up to 10 % from the given nominal values; these deviations are accounted for when performing transfer matrix simulations.

| Layer | Thickness (nm) | | | | | |
|---|---|---|---|---|---|---|
| | P-A | P-B | P-C | P-D | P-E | P-F |
| Ag | 100 | 100 | 100 | 100 | 100 | 100 |
| BPhen:Cs | **50** | **60** | **70** | **80** | **60** | **60** |
| BAlq | 10 | 10 | 10 | 10 | 10 | 10 |
| NPB:Ir(MDQ)$_2$(acac) | 40 | 40 | 40 | 40 | 40 | 40 |
| NPB | 10 | 10 | 10 | 10 | 10 | 10 |
| SpiroTTB:F$_6$TCNNQ | **105** | **95** | **85** | **75** | **105** | **105** |
| **SpiroTTB:Cl$_6$SubPc** | **24** | **24** | **24** | **24** | **36** | **48** |
| SpiroTTB:F$_6$TCNNQ | **60** | **75** | **90** | **105** | **60** | **60** |
| Ag | 25 | 25 | 25 | 25 | 25 | 25 |
| Substrate | 1.1E6 | 1.1E6 | 1.1E6 | 1.1E6 | 1.1E6 | 1.1E6 |

**Table S2. Device structure and layer thicknesses of red polaritonic OLEDs.** Nominal layer thicknesses for red polaritonic OLEDs comprising the emitter Ir(MDQ)$_2$(acac) and the strong coupling layer Cl$_6$SubPc with variations in cavity thickness (by changing the thickness of the SpiroTTB:F$_6$TCNNQ HTL and the BPhen:Cs ELT) and coupling strength (by changing the thickness of the SpiroTTB:Cl$_6$SubPc assistant strong coupling layer). Due to tooling variations and chamber positioning, the real film thicknesses can differ by up to 10 % from the given nominal values; these deviations are accounted for when performing transfer matrix simulations.

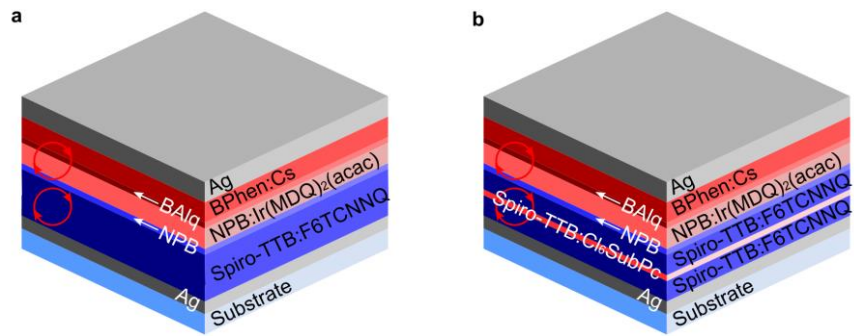

**Figure S1: Device architectures.** Schematic structures of the red reference (**a**) and polaritonic (**b**) OLEDs.

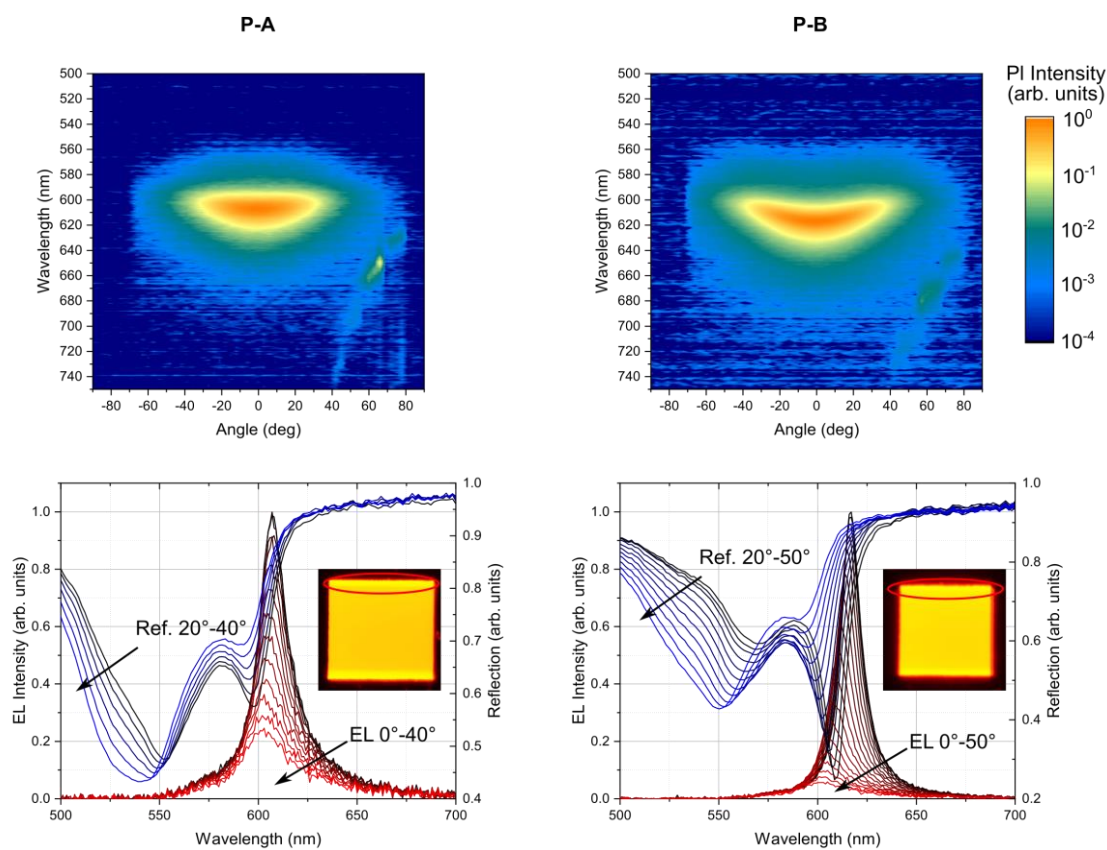

**Figure S2.** Angle-resolved electroluminescence of devices P-A and P-B on logarithmic colour scale, to show the weak shoulders in the emission spectra more clearly (top row). Electroluminescence (EL) and reflection (Ref.) spectra of the devices as individual lines, showing how emission from the shoulder at ~580 nm is angle-independent and in particular does not follow the UPB dispersion visible in the reflection spectra (bottom). This shoulder likely originates from emission outside the cavity defined by the anode-cathode overlap. The insets show photographs of the emitting POLEDs with a more blue-shifted emission coming from the top and bottom of the pixel (marked by red ellipse).

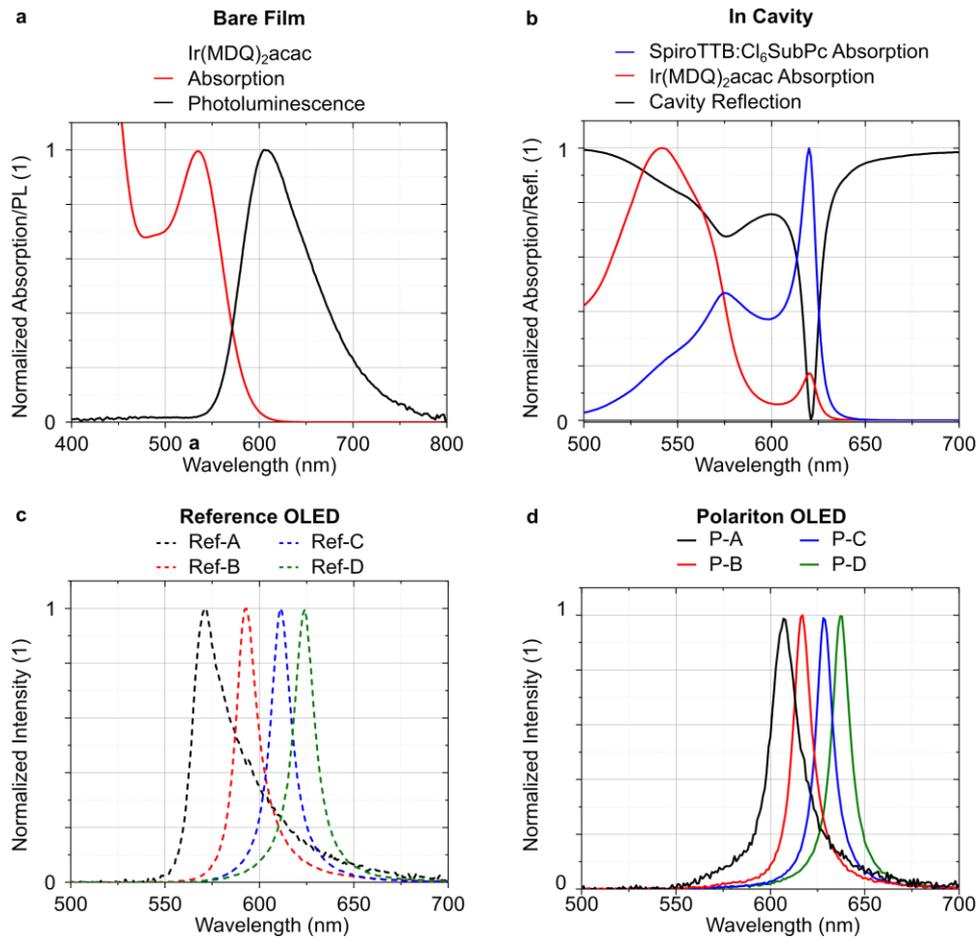

**Figure S3: Absorption of Ir(MDQ)$_2$(acac) and emission from lower polariton branch. a**, Normalized absorption and photoluminescence of a bare film of Ir(MDQ)$_2$(acac) doped at 10wt% into NPB. **b**, Simulated intra-cavity absorption spectra of the Ir(MDQ)$_2$(acac)-based EML, the assistant strong coupling layer of SpiroTTB-Cl$_6$SubPc, and spectral reflectivity of the cavity. The intra-cavity absorption of both the EML and the assistant strong coupling layer show the peak splitting characteristic of polariton-formation. The intra-cavity absorption of individual layers in the cavity was calculated by evaluating the electric field at every wavelength and position in the cavity via a transfer-matrix-method. Spectra are normalized to resolve the weak absorption of Ir(MDQ)$_2$(acac) in the presence of the stronger Cl$_6$SubPc absorption. Depending on exact device configuration and layer thicknesses, the absolute absorption of the Ir(MDQ)$_2$(acac)-based emission layer (EML) reaches between 1% and 5% of the absorption of the SpiroTTB:Cl$_6$SubPc assistant strong coupling layer. **c,d**, Emission spectra of reference OLEDs (**c**) and polariton OLEDs (**d**) at 0° incidence.

**Supporting Note 1:**

To quantify the contribution of the EML to strong coupling, we prepared bare films of NPB:Ir(MDQ)$_2$(acac) similar to the EMLs used in our POLEDs. This allowed us to quantify the absorption of the EML using VASE and showed that the absorption of the NPB:Ir(MDQ)$_2$(acac) EML is about 50x smaller than the absorption of the assistant strong coupling layers we use. Nevertheless, the absorption spectra show that there is well-defined absorption at the wavelength where strong coupling occurs (see Fig. S3 a). In an effort to untangle EML absorption from absorption in the strong coupling layer, we performed transfer matrix simulations with these new data to evaluate the intra-cavity electric field at every position and wavelength inside the devices. In doing so, we can calculate, as a function of wavelength, which fraction of an incident light wave is absorbed by each individual layer (compare e.g. Burkhard et al. [1]). In Fig. S3 b, we plot the calculated absorption of both the NPB:Ir(MDQ)$_2$(acac) EML and the SpiroTTB:Cl6SubPc assistant strong coupling layer for OLED P-B from our original manuscript. The absorption profiles of both layers show a splitting into two distinct polariton branches, which confirms our original interpretation that excitons in the EML of our devices emit through the lower polariton branch (LPB). The emission spectra of the polaritonic OLEDs show exclusive emission from the LPB, as shown in the comparative spectra in Fig. S3 c and d, and no outcoupling through the UPB is observed.

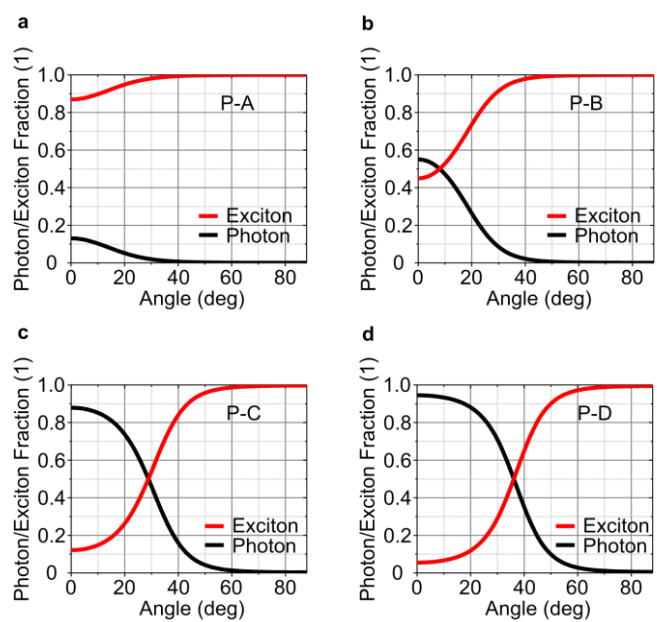

**Figure S4: Photon and exciton fractions of the different red polaritonic OLEDs. a-d**, Fraction of photon/exciton in the polariton for lower polariton branch (LPB) for devices P-A to P-D as function of angle. For the most blue-shifted cavities (P-A, P-B), the lower polariton exhibits a strongly excitonic behaviour resulting in a very flat dispersion. (In turn, the upper polariton in these cavities is strongly photonic, not shown.) For thicker cavities (P-C, P-D) the LPB starts off with a more photonic dispersion at small angles and turns excitonic only at larger angles.

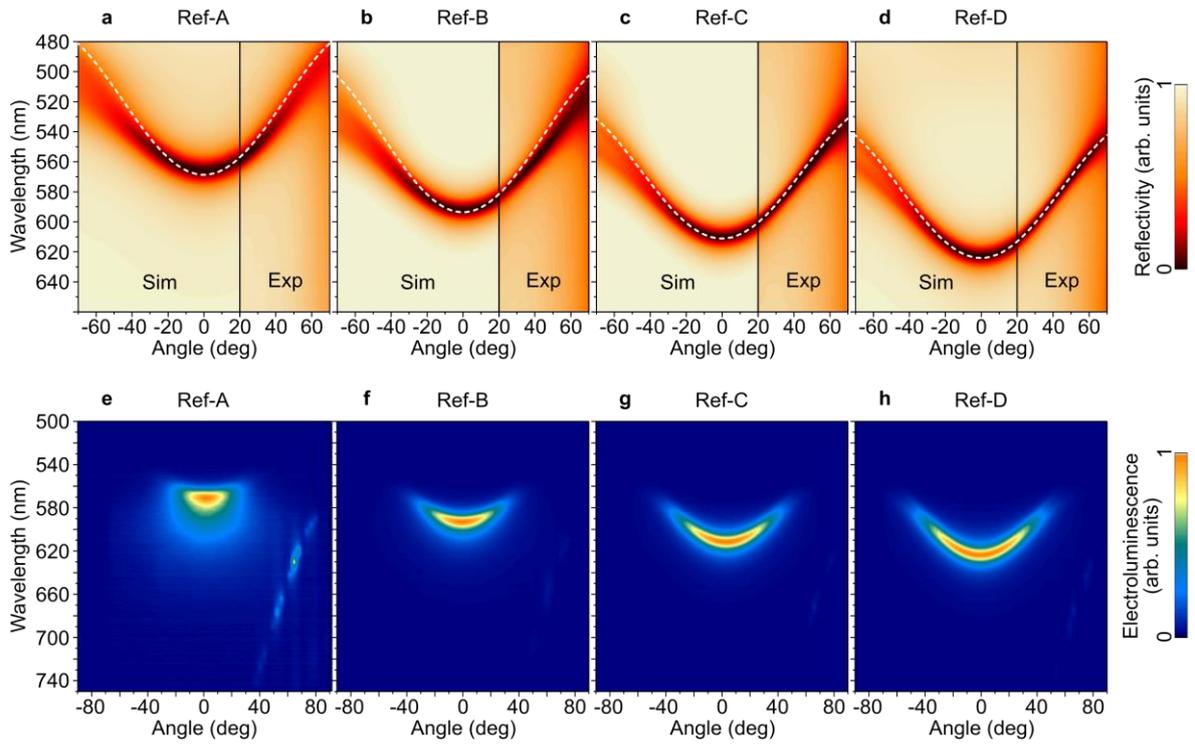

**Figure S5: Angle-resolved reflectivity and electroluminescence measurements of red reference OLEDs. a-d**, Simulated (Sim, left side) and measured (Exp, right side) reflectivity spectra of weakly coupled reference OLEDs in false colour scale with increasing cavity thickness. **e-h**, Measured angle-resolved electroluminescence of reference OLEDs corresponding to **a-d**. The additional emission visible between angles of 40 ° and 80 ° and wavelengths of 600 - 700 nm originates from light scattering at the substrate edge and should be disregarded as an artifact.

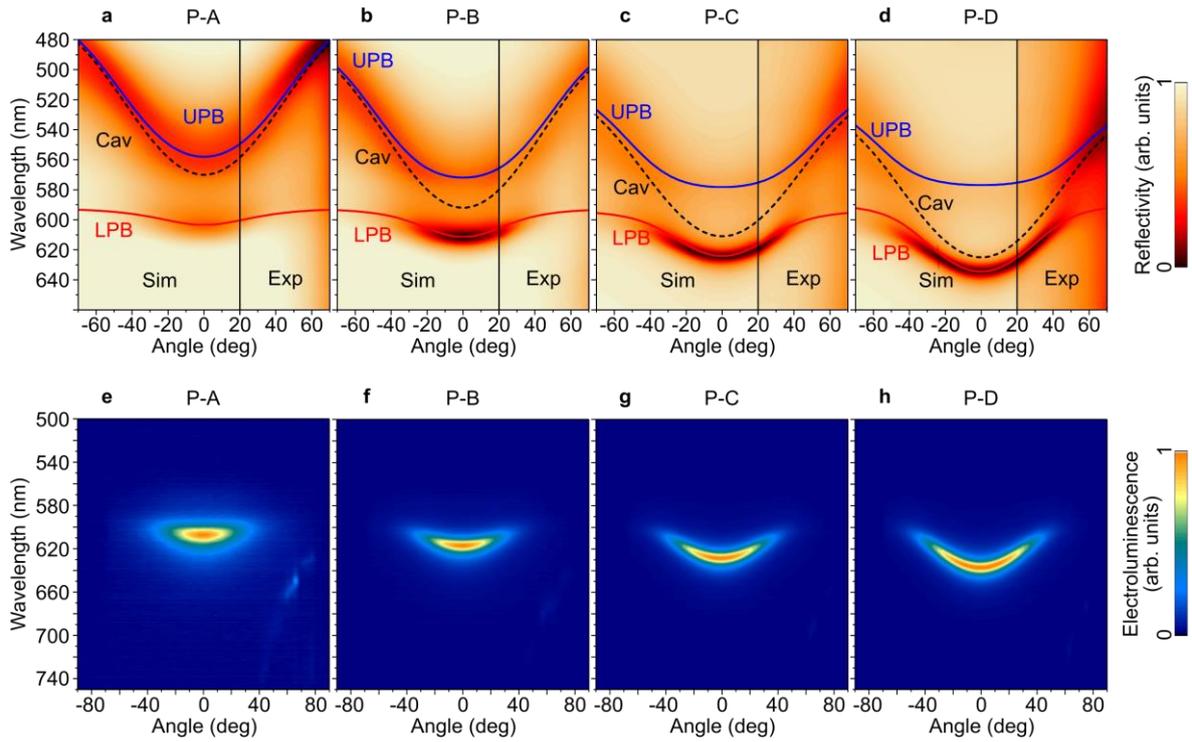

**Figure S6: Angle-resolved reflectivity and electroluminescence measurements of red polaritonic OLEDs. a-d**, Simulated (Sim, left side) and measured (Exp, right side) reflectivity spectra of strongly coupled polaritonic OLEDs in false colour scale. Solid lines show the position of the polariton modes according to the coupled oscillator model. **e-h**, Measured angle-resolved electroluminescence of polaritonic OLEDs corresponding to **a-d**. The additional emission visible between angles of 40 ° and 80 ° and wavelengths of 650 – 700 nm originates from scattered light at the substrate edge and should be disregarded as an artifact.

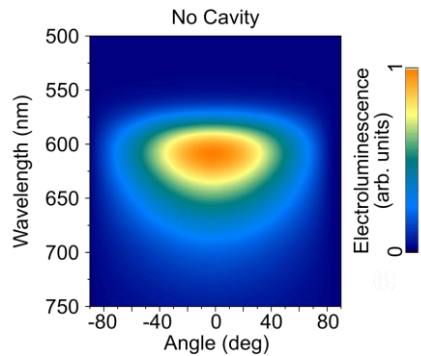

**Figure S7: Angle-resolved electroluminescence of a non-cavity red OLED.** Angle-resolved electroluminescence of a non-cavity bottom-emitting OLED with the structure: Substrate | ITO | Spiro-TTB:F6TCNNQ | NPB | NPB:Ir(MDQ)$_2$(acac) | BAlq | BPhen:Cs | Al. The transparent ITO bottom contact does not lead to a strong microcavity and thus the OLED shows broad (FWHM>70 nm), Lambertian emission.

| Layer | Thickness (nm) | |
|---|---|---|
| | P-Flexible | Ref-Flexible |
| Parylene-C | 5000 | 5000 |
| Nanolaminate | 54 | 54 |
| Parylene-C | 5000 | 5000 |
| Nanolaminate | 54 | 54 |
| Ag | 25 | 25 |
| Al | 1 | 1 |
| NBPhen:Cs | 60 | 60 |
| BAlq | 10 | 10 |
| NPB:Ir(MDQ)$_2$(acac) | 40 | 40 |
| NPB | 10 | 10 |
| SpiroTTB:F$_6$TCNNQ | **110** | **207** |
| **SpiroTTB:Cl$_6$SubPc** | **24** | **0** |
| SpiroTTB:F$_6$TCNNQ | **45** | **/** |
| MoO$_3$ | 1 | 1 |
| Ag | 80 | 80 |
| Al | 20 | 20 |
| Nanolaminate | 54 | 54 |
| Parylene-C | 3000 | 3000 |
| Nanolaminate | 54 | 54 |
| Parylene-C | 15000 | 15000 |

**Table S3. Device structure and layer thicknesses of red flexible reference and polaritonic OLEDs.** Nominal layer thicknesses for red flexible OLEDs with and without an assistant strong coupling layer of SpiroTTB:Cl$_6$SubPc. The individual layer thicknesses have been adjusted to keep the strong coupling layer and the emissive layer in their respective field maxima. Due to tooling variations and chamber positioning, real thicknesses can differ by up to 10 % from the nominal values given here; this is accounted for when performing transfer matrix simulations. The thin-film encapsulation (TFE) protects the device from oxygen and moisture and acts as quasi-substrate, as no other supporting substrate is employed. The TFE comprises micron-scale thick layers of parylene-C fabricated by chemical vapour deposition and nanolaminate films made of 9 pairs of 3 nm Al$_2$O$_3$ and 3 nm ZrO$_2$ (total 54 nm) fabricated by atomic layer deposition at 80 °C.

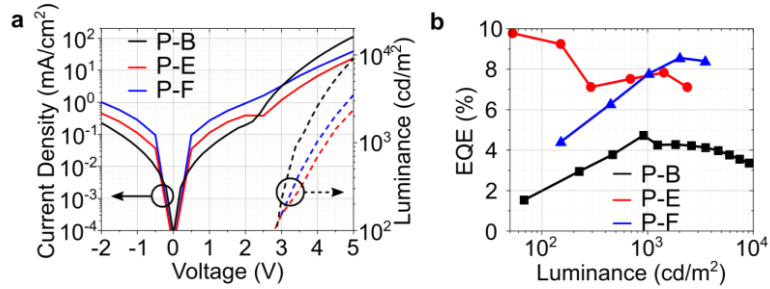

**Figure S8: OLED device performance for red polaritonic OLEDs with increasing coupling strength. a**, Current density and luminance over voltage for polaritonic OLEDs with increasing coupling strength. For thick strong coupling layers of SpiroTTB:Cl$_6$SubPc (i.e. P-E and P-F), the current density and luminance tend to be reduced, likely due to a charge imbalance introduced by the increasingly thick and electrically undoped strong coupling layer. **b**, External quantum efficiency (EQE) versus luminance for the same devices. Despite the reduced current densities for thicker strong coupling layers, the devices still exhibit similar EQE as the POLEDs described in the main text, with values reaching approximately 70 % of the EQE of comparable weakly coupled OLEDs. All EQE values have been corrected for any non-Lambertian emission.

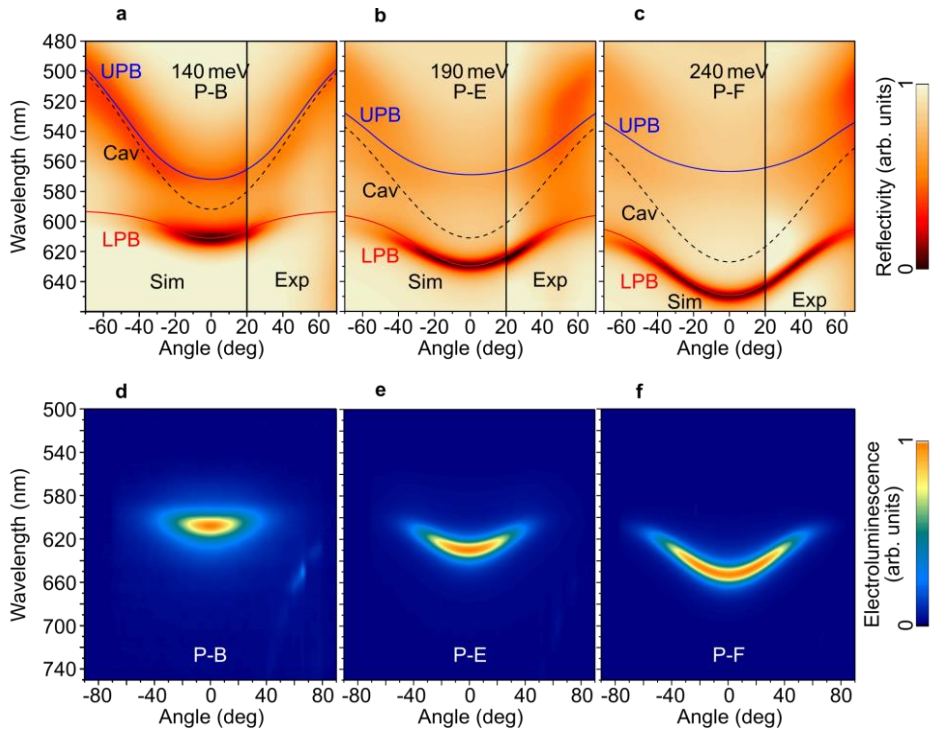

**Figure S9: Angle-resolved reflectivity and electroluminescence measurements of red polaritonic OLEDs with increasing coupling strength. a-c**, Simulated (Sim, left side) and measured (Exp, right side) reflectivity spectra of strongly coupled polaritonic OLEDs in false colour scale. Dashed black lines indicate the calculated position of the cavity mode in each device. Solid lines show the position of the polariton modes according to the coupled oscillator model. The value in meV in each panel indicates the Rabi-splitting energy derived from the coupled oscillator model. **d-f**, Measured angle-resolved electroluminescence of polaritonic OLEDs corresponding to (a)-(c). The stack architecture of the devices shown is given in Table S2.

| Layer | Thickness (nm) | |
|---|---|---|
| | Ref-G-C | Ref-G-D |
| Ag | 100 | 100 |
| BPhen:Cs | 40 | 40 |
| BPhen | 10 | 10 |
| Host:Ir(ppy)$_2$(acac) | 30 | 30 |
| TAPC | 10 | 10 |
| SpiroTTB:F$_6$TCNNQ | **195** | **215** |
| Ag | 25 | 25 |
| Substrate | 1.1E6 | 1.1E6 |

**Table S4. Device structure and layer thicknesses of green reference OLEDs.** Nominal layer thicknesses for green reference OLEDs comprising the emitter Ir(ppy)$_2$(acac) doped at 10 wt% into a host matrix of TCTA:BPhen (1:1). The cavity thickness is varied by changing only the thickness of the SpiroTTB:F$_6$TCNNQ HTL. Due to tooling variations and chamber positioning, the real film thicknesses can differ by up to 10 % from the given nominal values; this is accounted for when performing transfer matrix simulations.

| Layer | Thickness (nm) | | | | | |
|---|---|---|---|---|---|---|
| | P-G-A | P-G-B | P-G-C | P-G-D | P-G-E | P-G-F |
| Ag | 100 | 100 | 100 | 100 | 100 | 100 |
| BPhen:Cs | 40 | 40 | 40 | 40 | 40 | 40 |
| Bphen | 10 | 10 | 10 | 10 | 10 | 10 |
| Host:Ir(ppy)$_2$(acac) | 30 | 30 | 30 | 30 | 30 | 30 |
| TAPC | 10 | 10 | 10 | 10 | 10 | 10 |
| SpiroTTB:F$_6$TCNNQ | 90 | 90 | 90 | 90 | 90 | 90 |
| **C545T** | **18** | **18** | **18** | **18** | **12** | **24** |
| SpiroTTB:F$_6$TCNNQ | **40** | **60** | **80** | **100** | **60** | **60** |
| Ag | 25 | 25 | 25 | 25 | 25 | 25 |
| Substrate | 1.1E6 | 1.1E6 | 1.1E6 | 1.1E6 | 1.1E6 | 1.1E6 |

**Table S5. Device structure and layer thicknesses of green polaritonic OLEDs.** Nominal layer thicknesses for green polaritonic OLEDs comprising the emitter Ir(ppy)$_2$(acac) doped at 10 wt% into a host matrix of TCTA:BPhen (1:1) and the strong coupling layer C545T. By varying the lower HTL (SpiroTTB:F$_6$TCNNQ) and the strong coupling layer (C545T), we tune both the cavity mode position and the coupling strength. Due to tooling variations and chamber positioning, the real film thicknesses can differ by up to 10 % from the given nominal values; this is accounted for when performing transfer matrix simulations.

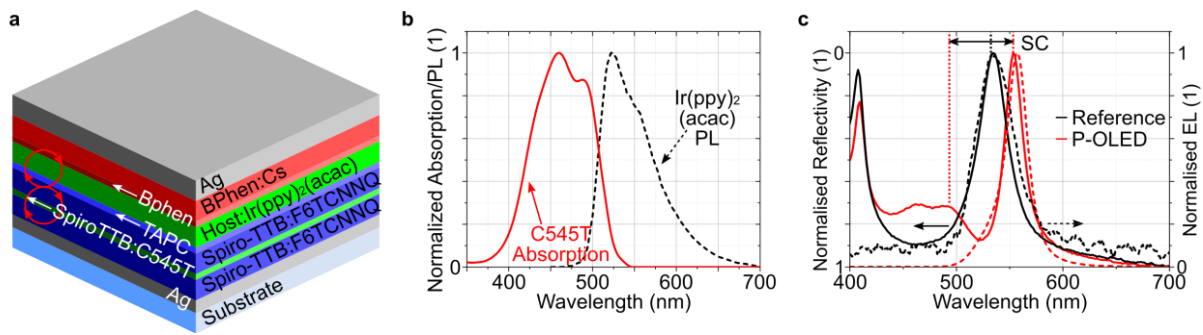

**Figure S10. Device and material properties for green polaritonic OLEDs. a**, Schematic of OLED structure comprising two reflective Ag contacts, Spiro-TTB doped with $F_6$TCNNQ as hole transport layer (HTL), TAPC as electron blocking layer (EBL), BPhen:TCTA (1:1) as host doped with Ir(ppy)$_2$(acac) as emissive layer (EML), BPhen as hole blocking layer (HBL), and BPhen doped with Cs as electron transport layer (ETL). To facilitate strong coupling (SC), an 18 nm layer of C545T is inserted in the HTL, at the peak of the optical field strength. (b) Absorption spectrum of C545T (red solid line) and photoluminescence (PL) spectrum of Ir(ppy)$_2$(acac) (black dashed line). (c) Left axis with inverted direction: Reflectivity spectra at 20 ° incidence of a reference microcavity OLED without the SC layer (black solid line) and a polaritonic OLED (POLED, red solid line). Right axis: Electroluminescence (EL) spectra recorded at an observation angle of 20 ° (dashed lines). In the POLED spectrum, the coupling of cavity photon and C545T exciton leads to a splitting of the resonances as indicated by the horizontal double arrow. EL is observed from the cavity mode in the reference OLED while it originates exclusively from the lower polariton in the POLED.

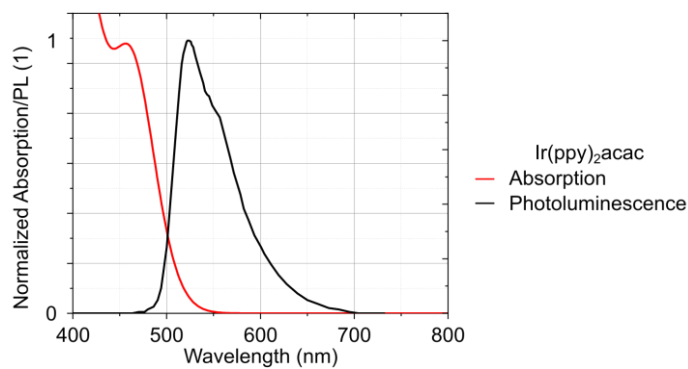

**Figure S11 Absorption and photoluminescence of Ir(ppy)$_2$(acac).** Absorption and photoluminescence spectra of a bare film of Ir(ppy)$_2$(acac) doped at 10wt% into a mixed host of TCTA:BPhen (1:1).

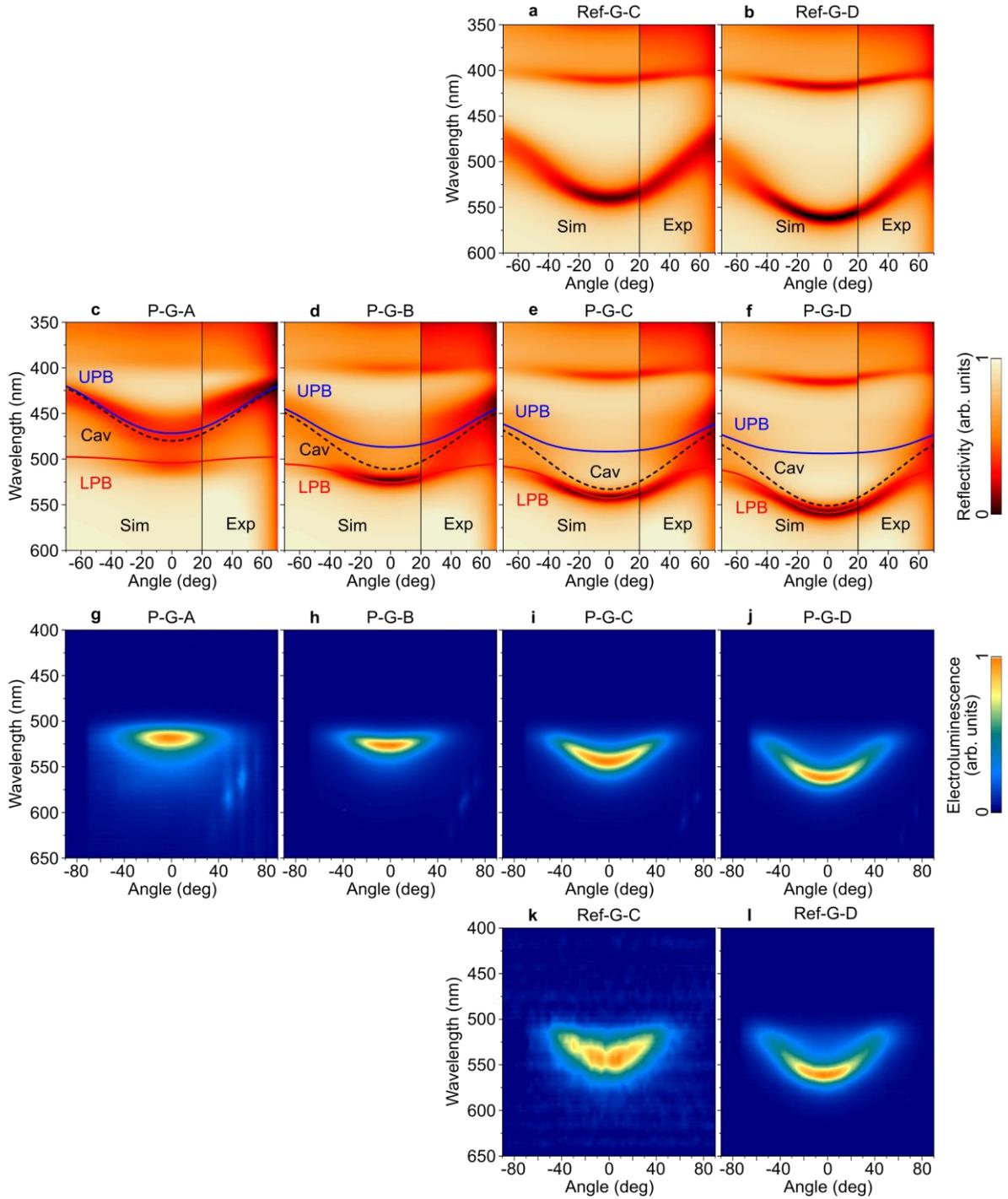

**Figure S12: Angle-resolved reflectivity and electroluminescence of green reference and polaritonic OLEDs. a-f**, Simulated (Sim, left side) and measured (Exp, right side) reflectivity spectra of green reference OLEDs (**a**, **b**) and polaritonic OLEDs (**c-f**) on a false colour scale. Solid lines show the position of the polariton modes according to the coupled oscillator model for a coupling strength of 180 meV. **g-l**, Measured angle-resolved electroluminescence of green polaritonic OLEDs (**g-j** corresponding to **c-f**) and reference OLEDs (**k**,**l** corresponding to **a**,**b**). The additional emission visible between angles of 40 ° and 80 ° and wavelengths of 550 - 650 nm originates from scattered light at the substrate edge and should be disregarded as an artifact. Reference OLEDs Ref-G-A and Ref-G-B, corresponding to P-G-A and P-G-B, respectively, showed poor performance due to unfavorable outcoupling efficiencies and are thus excluded from this figure.

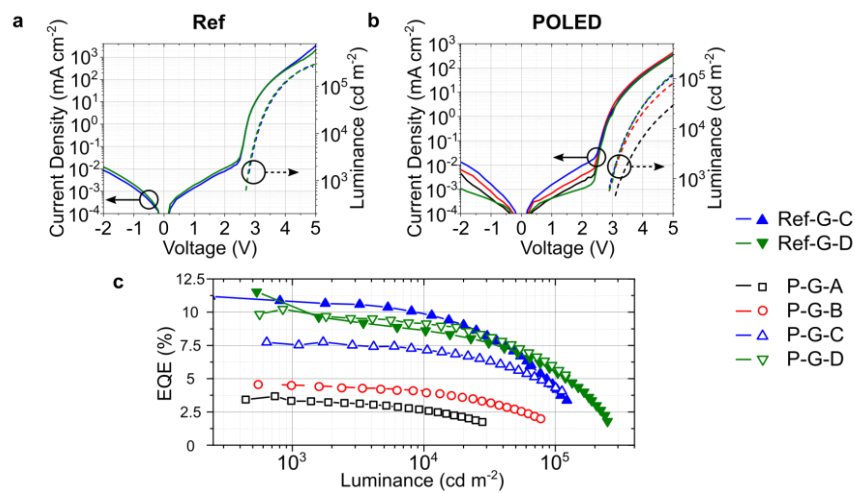

**Figure S13. Performance of green reference and polaritonic OLEDs. a,b**, Current density and luminance over driving voltage of the green reference MC OLEDs (**a**) and the POLEDs (**b**). All devices show a similar turn-on voltage regardless of whether the SC layer is present. The POLEDs reach a luminance of more than 100,000 cd/m² at 5 V. **c**, External quantum efficiency of the MC OLEDs (full symbols) and POLEDs (open symbols). POLEDs reach efficiencies of 75 % (P-G-C) or even 100 % (P-G-D) of their respective reference device.

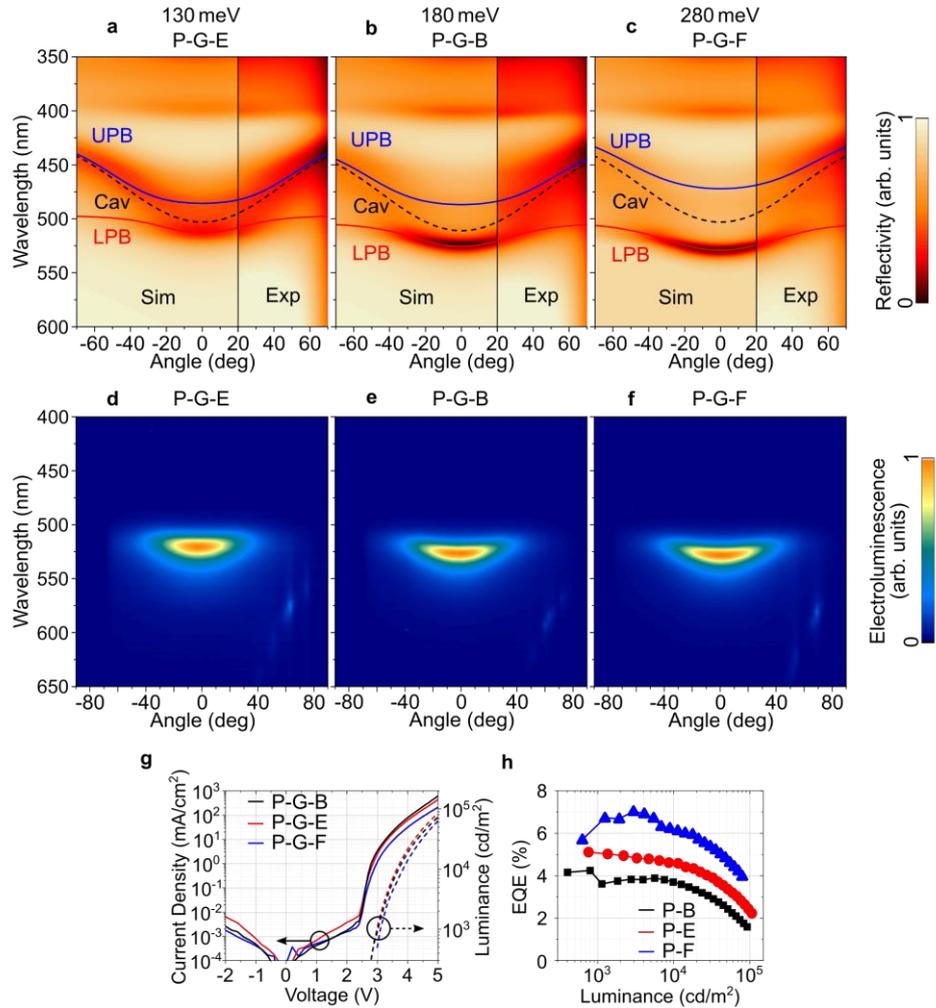

**Figure S14: Angle-resolved reflectivity and electroluminescence of green polaritonic OLEDs with increasing coupling strength. a-c**, Simulated (Sim, left side) and measured (Exp, right side) reflectivity spectra of strongly coupled polaritonic OLEDs in false colour scale. **d-f**, Measured angle-resolved electroluminescence of the polaritonic OLEDs in **a-c**. By increasing the thickness of the C545T SC layer, the Rabi-splitting energy can be tuned from 130 meV (P-G-E at 12 nm) to 180 meV (P-G-B at 18 nm) and to 280 meV (P-G-F at 24 nm). The additional emission visible between angles of 40 ° and 80 ° and wavelengths of 550 - 650 nm originates from scattered light at the substrate edge and should be disregarded as an artifact. **g**, Current density and luminance over driving voltage of polaritonic OLEDs with increasing coupling strength. With higher thicknesses of the C545T strong coupling layer, the current density and luminance drop slightly for any given voltage, likely due to a charge imbalance introduced by adding the electrically undoped strong coupling layer. **h**, External quantum efficiency (EQE) of the same devices. The EQE values have been corrected for any non-Lambertian emission.

| Layer | Thickness (nm) | | |
|---|---|---|---|
| | USC 1 | USC 2 | USC 3 |
| Ag | 100 | 100 | 100 |
| BPhen:CsO | 70 | 60 | 80 |
| BPhen | 10 | 10 | 10 |
| Host:Ir(ppy)$_2$(acac) | 30 | 30 | 30 |
| TAPC | 10 | 10 | 10 |
| SpiroTTB:F$_6$TCNNQ | 40 | 40 | 40 |
| C545T | **60** | **60** | **60** |
| SpiroTTB:F$_6$TCNNQ | 60 | 60 | 60 |
| MoO$_3$ | 1 | 1 | 1 |
| Ag | 25 | 25 | 25 |
| Substrate | 1.1E6 | 1.1E6 | 1.1E6 |

**Table S6. Device structure and layer thicknesses of green USC POLEDs.** Nominal layer thicknesses for the optimized green USC POLED (USC 1) as well as the blueshifted (USC 2) and redshifted (USC 3) POLEDs, each comprising a 60 nm C545T assistant strong coupling layer. The cavity thickness is varied by changing only the thickness of the BPhen:CsO ETL. Due to tooling variations and chamber positioning, the real film thicknesses can differ by up to 10 % from the given nominal values; this is accounted for when performing transfer matrix simulations.

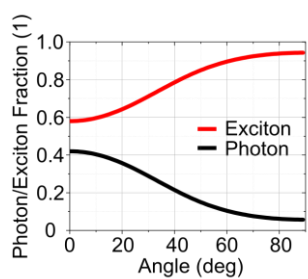

**Figure S15. Photon and exciton fractions of the green USC POLED.** Fraction of photon/exciton in the polariton for the lower polariton branch (LPB) of the USC POLED device also shown in Fig. 6 of the main text, revealing a significant excitonic fraction over all angles.

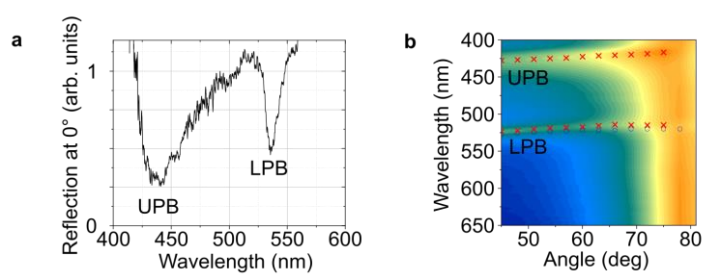

**Figure S16. Reflection spectra of the green USC POLED. a**, White-light reflection of the USC POLED at 0° measured by confocal micro-spectroscopy. **b**, Reflection spectra of the USC POLED between 45°-81° measured by variable angle spectroscopic ellipsometry. The reflection minima from **a** and **b** (red crosses) are plotted as crosses in Fig. 6 of the main text.

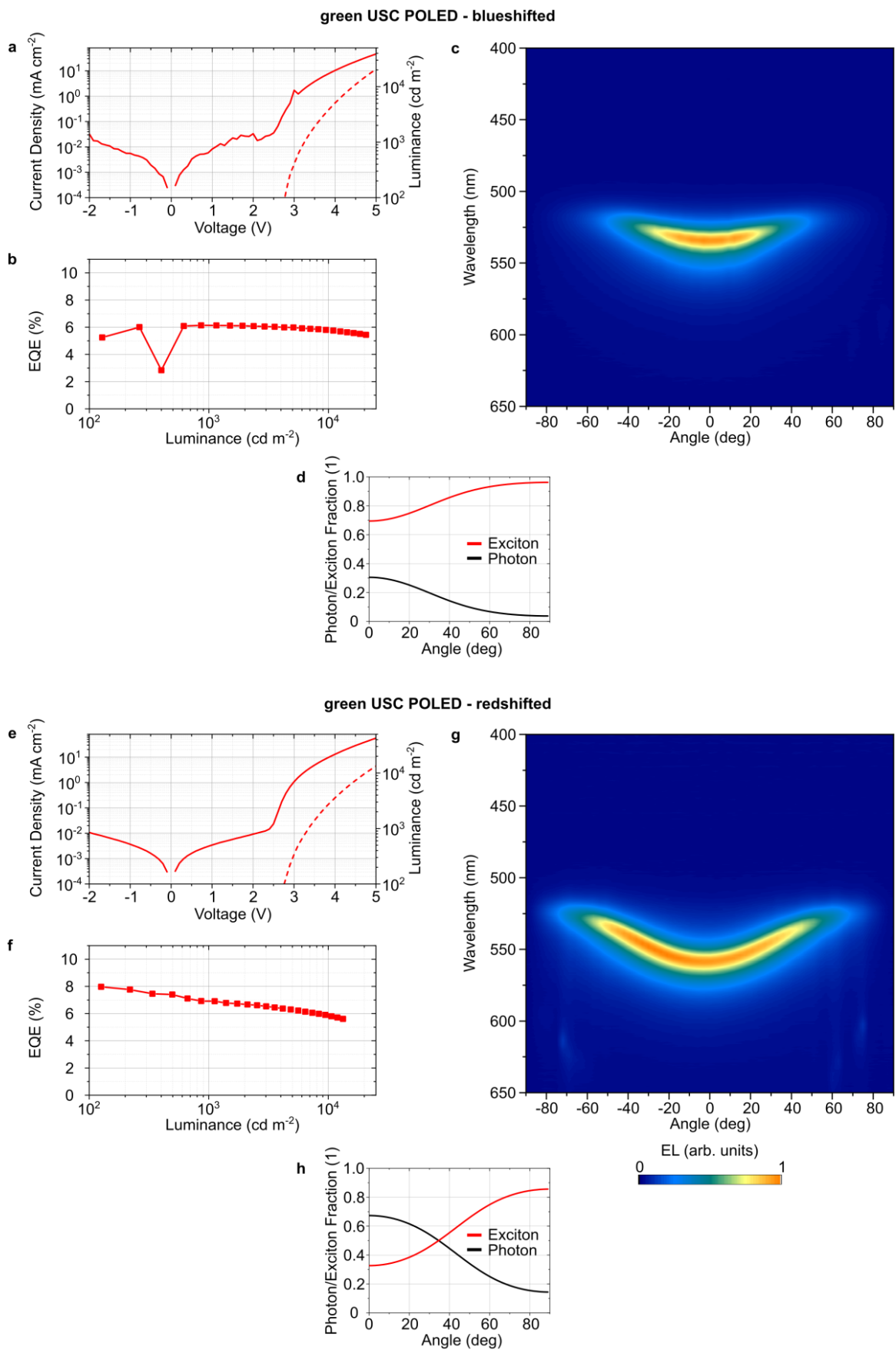

**Figure S17. Green POLEDs in the ultra-strong coupling regime at shifted emission wavelengths. a-c**, Blue-shifted USC POLED. **a,** Current density-voltage-luminance characteristics of the blue-shifted

device reaching a luminance of ~18,000 cd m$^{-2}$ at 5 V. **b**, External quantum efficiency of the blue-shifted device, reaching a peak of 6.1% at ~1,000 cd m$^{-2}$. **c,** Angle-resolved, measured electroluminescence (EL) intensity of the blue-shifted USC POLED, peaking at 534 nm. **a-d** Blueshifted USC POLED. **a** Current-density-voltage-luminance characteristics of the blue-shifted device reaching a luminance of ~18,000 cd m$^{-2}$ at 5 V. **b** External quantum efficiency of the blue-shifted device, reaching a peak of 6.1% at ~1,000 cd m$^{-2}$. **c** Angle-resolved, measured electroluminescence (EL) intensity of the blue-shifted USC-POLED, peaking at 534 nm. **d**, Exciton/Photon fractions of the blue-shifted device. **e-h**, Red-shifted USC POLED. **e**, Current density-voltage-luminance characteristics of the red-shifted device reaching a luminance of ~15,000 cd m$^{-2}$ at 5 V. **f**, External quantum efficiency of the red-shifted device, reaching a peak of ~8.0% at ~150 cd m$^{-2}$. **g**, Angle-resolved, measured EL intensity of the red-shifted USC POLED, peaking at 559 nm. **e-h** Redshifted USC POLED. **e** Current-density-voltage-luminance characteristics of the red-shifted device reaching a luminance of ~15,000 cd m$^{-2}$ at 5 V. **f** External quantum efficiency of the red-shifted device, reaching a peak of ~8.0% at ~150 cd m$^{-2}$. **g** Angle-resolved, measured electroluminescence (EL) intensity of the red-shifted USC-POLED, peaking at 559 nm. **h**, Exciton/Photon fractions of the red-shifted device.

**Supporting Note 2:**

The efficiency of the USC POLEDs achieved here is significantly higher than previously shown. Typically, one expects polariton emission efficiency to scale with their photon fraction and thus with negative detuning of the cavity[2,3], indicating a swift conversion from generated excitons to emitted photons. On the other hand, OLED efficiency strongly depends on the outcoupling efficiency, i.e. how efficiently internally generated photons escape from the device. This is largely driven by thin-film interference and thus there is an optimal optical thickness of the multilayer stack forming the OLED that depends on the emission spectrum of the specific emitter used. For our devices, due to the use of the assistant strong coupling layer (here C545T), we are in a situation where coupling strength and photon/exciton fraction of the polaritons can be controlled somewhat independently from the outcoupling efficiency of the specific emitter (here Ir(ppy)$_2$(acac)), while still remaining in the strong coupling regime. We show that by proper detuning as well as by proper choice of coupling strength, emission efficiency, coupling strength and angular dispersion can be optimized for simultaneously. This is demonstrated well in the sets of green POLEDs: First, for P-G-A to P-G-D, the Rabi splitting is fixed at 180 meV while the cavity is continuously tuned to the red. In turn, the quantum efficiency increases from ~3.5% (P-G-A) to ~10% (P-G-D). For the set of P-G-E/P-G-B/P-G-F/USC1 POLED, the detuning is kept relatively similar (close to 0), but an efficiency increase from ~4% (P-G-E) to ~7% (P-G-F) and finally ~10% (USC 1) can be observed for an increase in Rabi-splitting from 130 meV (P-G-E) to 520 meV (USC 1), as the LPB moves towards a more favourable spectral outcoupling window.